\documentclass[%
 reprint,
 superscriptaddress,
 amsmath,amssymb,
 aps,
 prl,
 gensymb,
floatfix,
ulem,
nolongbibliography,
]{revtex4-2}
\usepackage{gensymb}% for degree
\usepackage{graphicx}% Include figure files
\usepackage{dcolumn}% Align table columns on decimal point
\usepackage{bm, soul}% bold math
\usepackage{hyperref}% add hypertext capabilities
\usepackage[dvipsnames]{xcolor}
\hypersetup{
    colorlinks = true,
    linkcolor = blue,
    anchorcolor = blue,
    citecolor = blue,
    filecolor = blue,
    urlcolor = black
}

\graphicspath{}
\usepackage{multirow}
\begin{document}

\preprint{APS/123-QED}

\title{Path differences between quasistatic and fatigue cracks in anisotropic media}

\author{Xinyuan Zhai}
\affiliation{%
IMSIA, CNRS,   EDF, ENSTA Paris,  Institut Polytechnique de Paris, 91120 Palaiseau, France
}%
\author{Thomas Corre}
\affiliation{%
IMSIA, CNRS,   EDF, ENSTA Paris,  Institut Polytechnique de Paris, 91120 Palaiseau, France
}%
\affiliation{%
Nantes Université, Ecole Centrale Nantes, CNRS, GeM, UMR 6183, F-44000 Nantes, France
}%
\author{Ataollah Mesgarnejad }
\affiliation{%
Physics Department and Center for Interdisciplinary Research on Complex Systems, Northeastern University, Boston, Massachusetts 02115, USA
}%
\author{Alain Karma}%
\affiliation{%
Physics Department and Center for Interdisciplinary Research on Complex Systems, Northeastern University, Boston, Massachusetts 02115, USA
}%
\author{V\'eronique Lazarus}
 \email[]{veronique.lazarus@ensta-paris.fr}
\affiliation{%
IMSIA, CNRS,   EDF, ENSTA Paris,  Institut Polytechnique de Paris, 91120 Palaiseau, France
}%

\date{\today}% It is always \today, today,

\begin{abstract}
The propagation path of quasistatic cracks under monotonic loading is known to be strongly influenced by the anisotropy of the fracture energy in crystalline solids or engineered materials with a regular microstructure. Such cracks generally follow directions close to minima of the fracture energy. Here we demonstrate both experimentally and computationally that fatigue cracks under cyclic loading follow dramatically different paths that are predominantly dictated by the symmetry of the loading with the microstructure playing a negligible or subdominant role.  
\end{abstract}

%\pacs{62.20.Mk, 46.50.+a, 46.15.Ðx}
\maketitle

% ---------------------
%\paragraph{\bf Introduction}

Over and above the fascination with the shapes it creates \cite{Hul99}, understanding how and when cracks form  is of broad importance \cite{Cot10}, for instance in the design of manufacturing processes \cite{NamParKo12,Laz17,NgoKasImb18}, in geophysics for a better understanding of earthquakes \cite{FliBhaDmoRic05, rubino2022intermittent}, in mechanical engineering with obvious societal issues of safety \cite{AND05,MealorII2023}.
Cracks or sharp defects are known to particularly affect the resistance to fracture of any material as they lead to stress concentration in their vicinity \cite{Wil52}.
The theoretical framework \cite{GolSal74, CotRic80, HakKar09} and computational tools to predict brittle crack paths in diverse settings, including complex three-dimensional geometries \cite{BouMarMau14, CheCamLaz15}, are by now well-developed.
However,
 what criterion should be used to predict crack paths is still not fundamentally understood in the  practical case of {\it fatigue} crack growth occurring progressively under a repetitive cyclic load.
{Such kind of solicitations are common for industrial components (e.g. turbines, rotors, bridges), as well as  in earth and life sciences (e.g. waves, tidal ebb and flow, weather or climate cycles, walking/running steps).}

To date, it has been commonly assumed that fatigue cracks follow the same path as {\it quasistatic} cracks that propagate infinitesimally slowly (\emph{i.e.,} at a velocity much smaller than the sound speed) under a monotonic, as opposed to {\it fatigue} cyclic load \cite{BucCheRic04, CheCamLaz15}. 
This assumption is based on the qualitative observation that fatigue and quasistatic cracks follow a similar path in {\it isotropic} materials, \emph{i.e.,} materials with both isotropic elastic properties and a fracture energy independent of the direction of the crack path. 
In such materials, the principal of local symmetry (PLS) has been traditionally used to predict quasistatic crack paths \cite{GolSal74} for 2D plane strain or plane stress within the traditional framework of LEFM; 3D out of plane mode III loading causes crack front segmentation, but the propagation of segmented fronts remains consistent with the PLS \cite{CheCamLaz15}. The PLS assumes that the stress distribution remains symmetrical about the instantaneous crack axis during propagation and corresponds to a vanishing mode II stress intensity factor ($K_{II}=0$). It stems from the fact that cracks with non-vanishing $K_{II}$ kink by a discontinuous change of propagation direction. Therefore, cracks following curvilinear paths, with a continuous change of direction, should propagate with $K_{II}=0$ \cite{CotRic80}. Such crack paths also maximize the energy release rate, \emph{i.e.,}, the rate at which the stored elastic energy in the material is released taking into account the energy cost $\Gamma$ (per area) of creating new fracture surfaces. The PLS and Maximum Energy Release Rate (MERR) criteria predict slightly different kink angles \cite{AmeLeb92} for a crack initially loaded with a finite $K_{II}$, but the same curvilinear paths with $K_{II}=0$.

For anisotropic materials, the PLS is not applicable but extensions of the MERR criterion have been used to predict crack paths in various contexts \cite{marder_cracks_2004,Hakim2005, HakKar09,BicRomTak13,MesPanErb20}.
A MERR criterion for anisotropic media was formally derived within the diffuse interface phase-field approach to brittle fracture \cite{Hakim2005,HakKar09} by taking into account the anisotropy of the fracture energy quantified by the function, $\Gamma(\theta)$, where $\theta$ is the angle between the instantaneous crack propagation direction and some fixed reference axis. The selected crack path was shown to be the one that maximizes with respect to $\theta$ the energy released $G(\theta)-\Gamma(\theta)$, where $G(\theta)$ quantifies the elastic strain energy available by unit surface to propagate the crack. Consequently, the crack is predicted to propagate with a finite $K_{II}$ proportional to $d\Gamma(\theta)/d\theta$ in anisotropic media, and thus $K_{II}=0$ in isotropic media consistent with the PLS. 

The MERR criterion has been validated for quasistatic brittle fracture by phase-field simulations \cite{HakKar09,MesPanErb20} as well as experiments using materials with a composite microstructure engineered to have nearly isotropic elastic properties but a strongly anisotropic fracture energy \cite{MesPanErb20}. However, it is not expected to apply to fatigue cracks that propagate under cyclic loads that peak at $G(\theta)$ magnitudes well below the Griffith threshold $\Gamma(\theta)$; fatigue cracks do not obey the energy balance condition $G(\theta)=\Gamma(\theta)$, which is valid at each instant of propagation of a quasistatic crack. Fatigue crack growth is more commonly described by a phenomenological Paris-type law $da/dN=C\Delta K_I^n$, which relates the crack advance per cycle to the variation $\Delta K_I$ of mode I stress intensity factor during each cycle. For fatigue crack growth in isotropic media, the PLS can be rationalized in the same way as for quasistatic fracture by remarking that cracks with $K_{II}\ne 0$ kink, and that therefore smooth cracks should propagate with $K_{II}=0$. However, there is no reason to expect that the MERR criterion with $K_{II}\sim d\Gamma(\theta)/d\theta$ is applicable to fatigue crack paths. Hence, it remains largely unknown how fatigue and quasistatic crack paths differ from each other in anisotropic media and, in the absence of a MERR criterion, what physically controls the former.

\begin{figure}[htbp!]
\includegraphics[width=\linewidth]{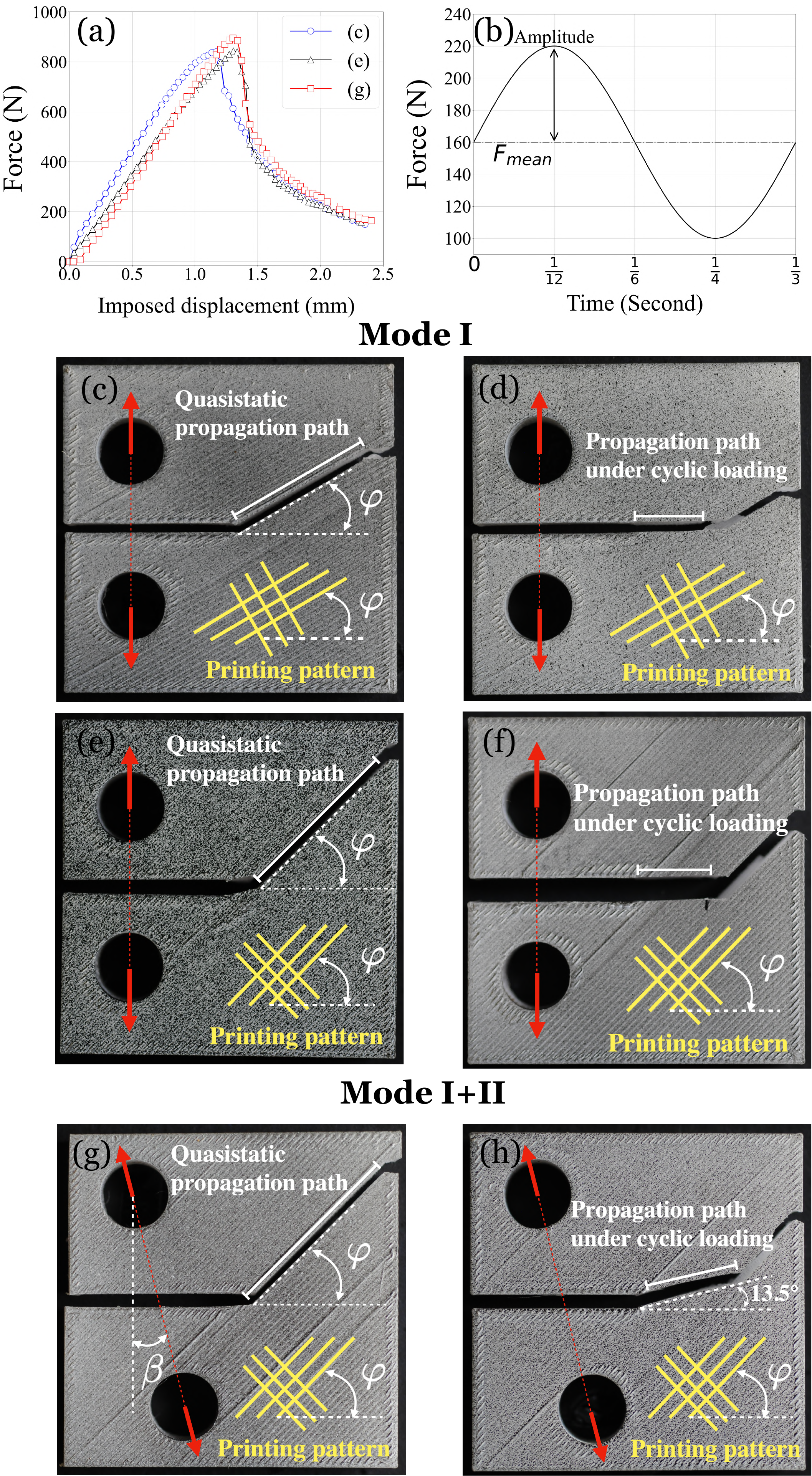}% Here is how to import EPS art
\caption{
\label{fig:paths} 
{Different crack propagation paths observed under monotonic load (left column) and cyclic load (right column) in polycarbonate CT (c-f) or CTS (g, h) samples printed by FDM. Under increasing imposed displacement (a), the crack propagates quasistatically in the weak plane of the microstructure (c, e, g) until brutal failure in another direction. Under cyclic load (b), propagation occurs along a different path  (d, f, h) during all the fatigue cycles. Bifurcation in the weak plane occurs an increasing load is applied to completely break the specimen. (Samples geometry: 50 $\times$ 48 $\times$ 8  mm$^3$. c,d:  $\varphi=30 \degree$; e, f: $\varphi=45 \degree$;g, h: $\varphi=45\degree$, $\beta=15 \degree $) }} 
\end{figure}

In this Letter, we address those questions by comparing quasistatic and fatigue crack paths both experimentally and computationally in two model materials where the elastic properties are isotropic but the fracture energy is strongly anisotropic. We demonstrate that in both cases quasistatic and fatigue cracks follow strikingly different paths. The experiments use 3D-printed samples with a criss-cross microstructure yielding a fracture energy $\Gamma(\theta)$ with a strong four-fold anisotropy. As illustrated in Fig.~\ref{fig:paths}, in samples with the same printed structure and loading in pure tensile mode I or with the addition of a shear mode II component, quasistatic cracks follow non-vanishing $K_{II}$ directions corresponding to sharp local minima of the fracture energy, as expected from MERR. In contrast, somewhat surprisingly, fatigue cracks follow vanishing $K_{II}$ directions despite the strong anisotropy of the underlying microstructure. The phase-field computations use a different two-fold anisotropic form of $\Gamma(\theta)$, which was previously used to validate the MERR criterion for quasistatic crack paths in anisotropic media \cite{HakKar09,MesPanErb20}. They yield similar results as the experiments with the difference that fatigue cracks follow paths where $K_{II}$ is much smaller than along quasistatic crack paths, but non-vanishing. We describe the experiments and computations in what follows and further interpret our findings.\\

%\paragraph{\bf Experiments}
% Printing strategy
To address the question experimentally,  polycarbonate Compact Tensile  (CT) and Compact Tensile Shear samples (CTS)  are printed by Fused {Deposition} Modeling (FDM) following a 100\% filling criss-cross pattern: each successive layer is oriented at 90$\degree$ to the previous one, with a $\varphi$ inclination to the initial crack Fig.~\ref{fig:paths}).
Tensile tests on samples with different filling angles, showed that the material is nearly isotropic with Young modulus $E\sim 2000$~MPa  and Poisson' ratio $\nu=0.34$ \cite{CorLaz21}. Once printed, a sharp crack is introduced by fatigue \cite{Zhai2023phd} and black-and-white speckle  is applied on the samples prior to testing for Digital Image Correlation (DIC).
%
%Loading
{To propagate the crack, the samples are loaded in a tensile machine either by increasing monotonically the displacement at a constant rate for the quasistatic propagation case or by applying a cyclic force load for fatigue} (Fig.~\ref{fig:paths}a,b). While the load is  mirror-symmetric to the crack plane for CT samples corresponding to pure mode I, a non-zero $\beta$ tilt for CTS samples induces non-symmetrical mode I+II {loading}.

Figure~\ref{fig:paths} highlights the difference between quasistatic and fatigue propagation on a set of representative cases. 
Applying increasing displacement, the force-displacement curve has a mountain shape (Fig.~\ref{fig:paths}a).  Before the peak, no crack propagation occurs and the force displacement curve is quasi-linear, which is consistent with linear elasticity. After the peak, the crack propagates quasi-statically and the force drops slowly. 
Fracture occurs in the $\varphi$ direction  (Fig.~\ref{fig:paths}c, e, g) for both CT and CTS. At the microstructure scale, it corresponds to separating two parallel threads every other layer, and to the orthogonal breaking of the threads on the remaining layers. 
Applying cyclic loading (Fig.~\ref{fig:paths}b), the crack progressively advances with each loading cycle. The crack advances straight ahead for CT samples (Fig.~\ref{fig:paths}d, f), while a kink is observed for CTS (Fig.~\ref{fig:paths}h) but in a different direction to that obtained under monotonic loading (Fig.~\ref{fig:paths}g). 
At the microstructure scale, the crack does not generally follow the orientation of the threads.
We have performed experiments for different values of  $\beta$ and $\varphi$, with good reproducibility and similar differences in the propagation paths. For each case, the crack is oriented by the microstructure for monotonic loading and by the loading orientation for fatigue.\\

%\paragraph{\bf Explanations} 

\begin{figure}[htbp!]
\includegraphics[width=\linewidth]{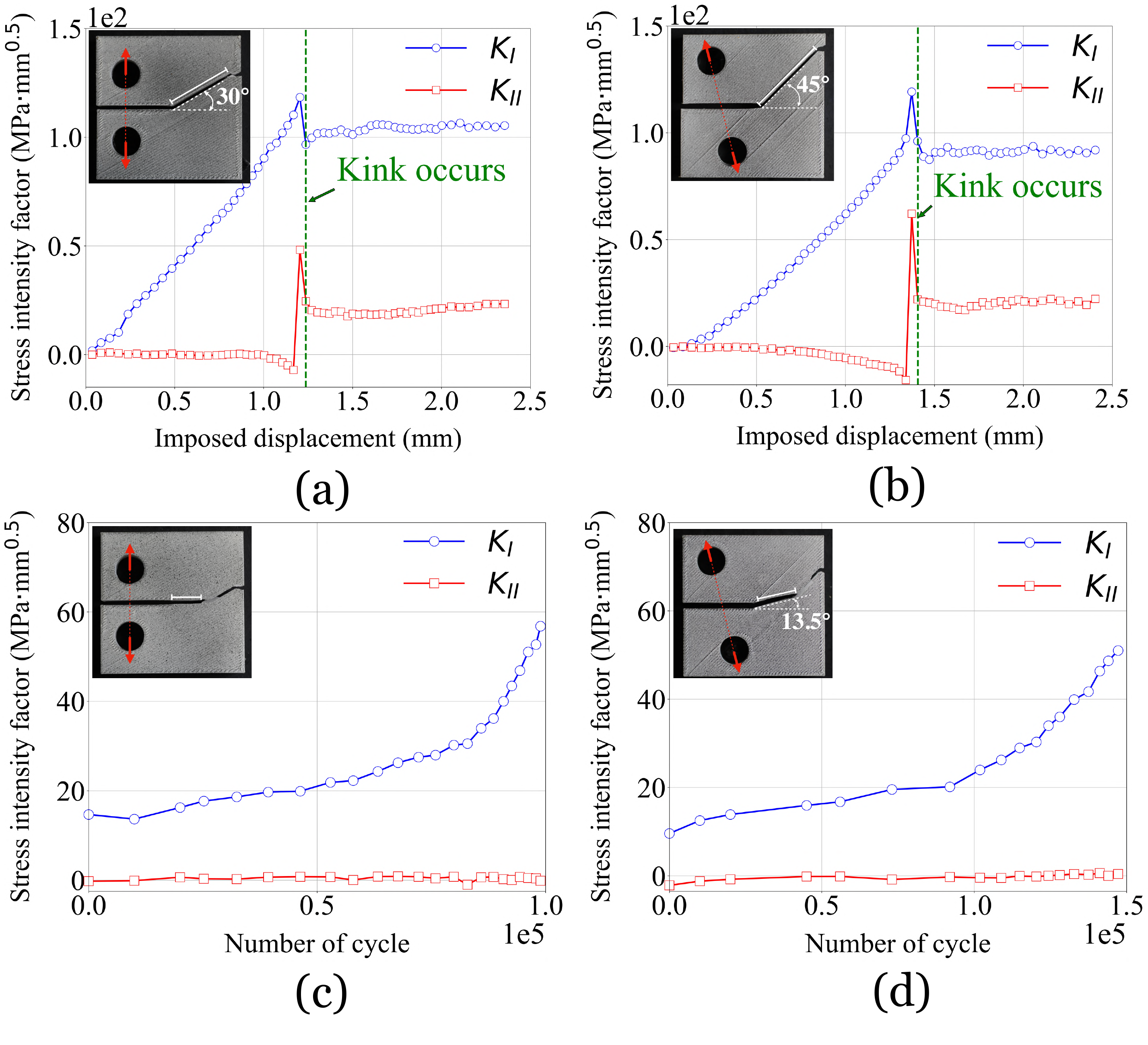}% Here is how to import EPS art
\caption{
\label{fig:SIF} Evolution of the stress intensity factors { during crack propagation measured by DIC. They are measured along the path indicated by a white segment in the insets. $K_{II} \neq 0$ during quasistatic propagation (a, b). $K_{II}=0$ in both CT and CTS during fatigue propagation (c, d)} }
\end{figure}
To find out more about the underlying crack propagation rules, and in particular if the PLS is retrieved in fatigue,  {the stress intensity factors ($K_{I}$, $K_{II}$) are measured  by DIC } \cite{Hamam2007, CorLaz21}
using the open-source software UFreckels \citep{Rethore2015,Rethore2018soft}. Figure \ref{fig:SIF} shows their evolution during crack propagation for the experiments of Fig.~\ref{fig:paths}~(c, d, g, h).
For monotonic load and CT samples (Fig.~\ref{fig:SIF}a), only mode I builds up before propagation while the applied load is increased;  for  CTS samples (Fig.~\ref{fig:SIF}b), mode I and II contributions increase. This is {due to} the symmetry and non-symmetry of the CT and CTS setups respectively. 
At the onset of propagation, the quasistatic crack changes direction and a jump in the value of $K_I$ and $K_{II}$ is observed. This jump is consistent with LEFM \cite{AmeLeb92} as demonstrated in  \cite{CorLaz21}. As $K_{II} \neq 0$, the PLS is not verified but the crack propagation direction can be predicted by a MERR criterion as in previous studies of quasistatic crack propagation in anisotropic media \cite{marder_cracks_2004,Hakim2005,HakKar09,BicRomTak13,MesPanErb20, Zhai2023phd, Suppinfo}.
For cyclic load (Fig.~\ref{fig:SIF}c, d), the crack propagates as soon as cycling begins. The stress intensity factors evolution is the same as {that expected} for an isotropic material: (i) $K_I$  increases for fixed applied force from the increase of the crack length \cite{IrwParTad00}, (ii) $K_{II}=0$ as soon as the crack propagates in agreement with the Principle of Local Symmetry. {In brief, contrary to the quasistatic case}, fatigue propagation follows the propagation rules of isotropic material and is not affected by the anisotropy of the microstructure.\\

To illustrate computationally the difference between quasistatic and fatigue crack paths in anisotropic media, 
we used the two-fold anisotropy \cite{HakKar09, MesPanErb20}
$\Gamma(\theta)=\Gamma_{max}\sqrt{{\cal A}^{-2}\cos^2(\theta-\phi)+\sin^2(\theta-\phi)}$
where ${\cal A}=\Gamma_{max}/\Gamma_0$ measures the toughness ratio for propagation perpendicular to ($\Gamma_{max}$) and parallel to ($\Gamma_0$) a weak plane analogous to one of the thread axis in the experiments. This form causes quasistatic cracks to follow directions close to weak planes for sufficiently strong anisotropy \cite{HakKar09, MesPanErb20}.
%, as opposed to being locked to weak planes for the form with cusps used to interpret the experiments. However, this difference is unimportant for the present purpose of illustrating the difference between quasistatic and fatigue crack paths in anisotropic media, irrespective of whether $\Gamma(\theta)$ contains cusps or not. 
Quasistatic propagation was modeled using the well-established alternate minimization scheme applied to anisotropic media in \cite{MesPanErb20}. Fatigue crack growth was investigated using a recently developed phase-field model of fatigue \cite{GroMesKar22} that enables crack growth below the Griffith threshold ($G<\Gamma$) by degradation of the fracture toughness \cite{MesImaKar19}. 
%\colVLbar{This formulation also treats the number of cycles $N$ as a continuous variable, thereby making it possible to describe crack paths on experimentally relevant time scales where $N\gg 1$.} 
This formulation has reproduced salient features of fatigue crack growth, such as the Paris law and non-trivial 2D and 3D crack paths in isotropic media \cite{GroMesKar22}. 
 We simulate  compact specimen of width $W$ \cite{ASTM:E-1820-05}. {Details of the numerical methods are given in the supplementary material \cite{Suppinfo}}. %\colVL{@Alain, @Ata. To be removed, make a supplemental material with numerical details? with dimensions $1.25W\times1.2W$ and a $W/20$ initial precrack. We set $\lambda/\mu=1$ (\emph{i.e.,} $E=8\lambda/3$, $\nu=1/3$) where plane-stress Lame parameters are $\lambda=E\nu/(1-\nu^2)$ and $\nu=E/(2(1+\nu))$  and set the phase field length scale  $\xi=W/100$.}
The loading is applied via a vertical displacement $u_y$ on the top half of the upper hole and by setting $u_x=u_y=0$ around the lower hole.
In this geometry, the energy release rate decreases for displacement-controlled loading as the crack propagates. Therefore, in both quasi-static and fatigue simulations, we set the displacement $u_{y}=k \ell$ where $\ell=2\sqrt{\Gamma_{max} W/3E}={3E}K_{IC}^{max}\sqrt{3W}/2$ and slowly  increase $k$ such that quasi-static crack propagation remains energetically stable and fatigue cracks are able to propagate an extended distance. 
For quasi-static propagation, we vary $k$ from 0 to 4. The onset of crack propagation occurs at $k\approx 1.6$ and $k\approx 2.6$ for $\phi=30^\circ$ and $\phi=90^\circ$, respectively, followed by quasi-static crack propagation when $k$ is increased above these thresholds.
For fatigue, we set $k=1.5$ for $\phi=90^\circ$ below its quasi-static limit. We increase $k$ from 1 to $2.25$ for $\phi=30^\circ$, which ensures that the energy release rate at the crack tip remains at all times above the minimum threshold $G_{min}$ for fatigue crack growth but below the Griffith threshold $G=\Gamma$ for quasi-static propagation. 
Further details of the phase-field methods can be found in \cite{MesPanErb20} for quasi-static and \cite{Grossman-Ponemon:2022wl} for fatigue crack propagation. %\colVLbar{Fatigue simulations were carried out using the parameters $e_{\min}/\lambda=2$, $\gamma_{\min}=0.025$, and by setting the time scale of toughness degradation $\tau=1$. Fatigue crack growth only depends on the dimensionless loading rate $\tau dk/dt$ that determines the value of $G_{min}<G<\Gamma$ at the crack tip with $G\approx G_{min}$ in the limit $\tau dk/dt\rightarrow 0$.}

\begin{figure}[htbp!]
\includegraphics[width=\linewidth]{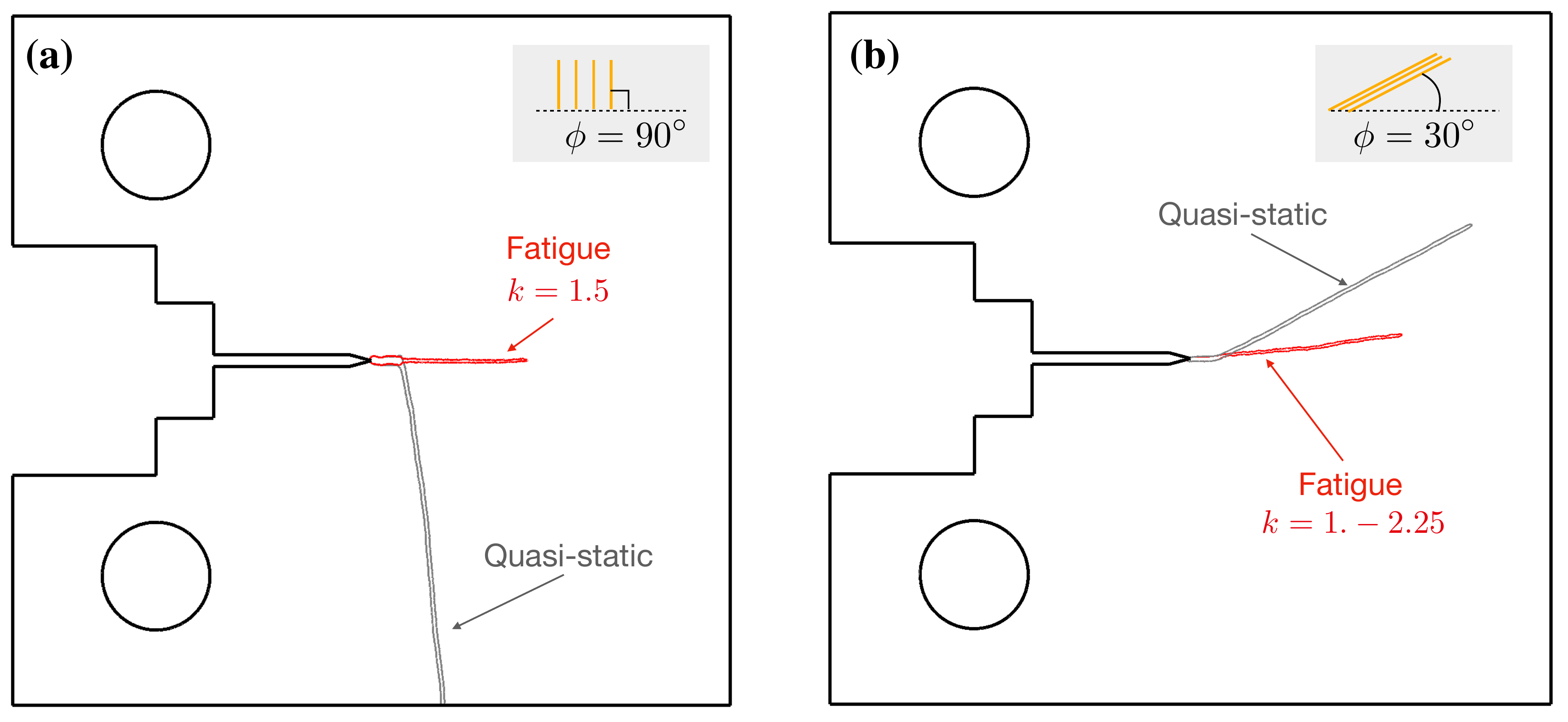}
\caption{\label{fig:PF} Comparison of computed quasi-static (gray) and fatigue (red) crack paths with the weak plane at (a) $\phi=90^\circ$ (b) $\phi=30^\circ$ with respect to the horizontal propagation axis. %($R_{fatigue}/\xi=100/9\pi$).
}
\end{figure}

The results are shown in Fig.~\ref{fig:PF} which compares quasistatic and fatigue crack paths for a large anisotropy $\mathcal{A}=4$ and two different weak plane orientations. When the weak plane is perpendicular to the horizontal axis (left panel of Fig.~\ref{fig:PF} for $\phi=90^\circ$), the quasistatic crack kinks at a large angle to follow a direction close to the weak plane with $K_{II}\ne 0$. In contrast, the fatigue crack propagates straight with $K_{II}=0$. When the weak plane is at a $30^\circ$ angle with respect to the horizontal axis (right panel of Fig.~\ref{fig:PF} for $\phi=30^\circ$), the quasistatic crack kinks to follow again a direction close to the weak plane, while the fatigue crack is much more weakly defected to propagate in a direction with $K_{II}$ much smaller than the quasistatic crack, but non-vanishing. \\

From a general symmetry viewpoint, we would expect fatigue cracks to follow a direction that is influenced both by the symmetry of the loading, which dictates a preferential axis for the cyclic damage process (\emph{e.g.,} the principal stress axis), and the symmetry of the underlying microstructure of the material, which offers the possibility for damage to occur preferentially in certain directions (\emph{e.g.,} parallel to the thread axis). 
The experiments demonstrate that the symmetry of the loading plays a dominant role for the printed materials investigated here. 
Under pure tensile loading, the stress fields are mirror-symmetric about the horizontal propagation axis and fatigue cracks propagate along this axis. They follow the PLS irrespective of the orientation of the microstructure (\emph{i.e.,} $\varphi=45^\circ$ and $\varphi=30^\circ$ yield identical fatigue crack paths with $K_{II}\approx 0$, {Fig.~\ref{fig:paths}d, f} ). In contrast, in the computations, the symmetry of the microstructure plays a subdominant, albeit not entirely negligible, role. For $\phi=90^\circ$, both the loading and the microstructure (\emph{i.e.,} $\Gamma(\theta)$) are mirror-symmetric about the horizontal axis and the fatigue crack propagates straight with $K_{II}=0$. For $\phi=30^\circ$, the microstructure is no longer mirror-symmetric about the horizontal axis, thereby causing the fatigue crack to propagate at a small angle. Further studies are warranted to elucidate the relative importance of loading and microstructure in dictating fatigue crack paths in different anisotropic materials. The present experiments and computations clearly demonstrate that loading plays a dominant role in two model cases where quasistatic crack paths are largely dictated by the microstructure.

The research at Northeastern University was supported by the United States NSF-CMMT grant 1827343.
The research at ENSTA Paris was partially supported by Agence de l'Innovation de Défense – 
AID - via Centre Interdisciplinaire d’Etudes pour la Défense et la 
Sécurité – CIEDS (AID-2018 60 0071 00 470 75 01 and FracAddi project 2022)
and by IP Paris Doctoral Allocation.  The authors thanks N. Thurieau, L. Cherfa for their help in the experiments.
\nocite{}

{For the purpose of Open Access, a CC-BY public copyright licence (\url{https://creativecommons.org/licenses/by/4.0/})
has been applied by the authors to the present document and will
be applied to all subsequent versions up to the Author Accepted
Manuscript arising from this submission. }
%\bibliography{../references}% Produces the bibliography via BibTeX.

\end{document}